\documentclass[12pt,letterpaper]{article}
\newcommand{\rom}[1]{\mathrm{#1}}

\newcommand{\p}{\partial}

\newcommand{\ep}{\epsilon}

\newcommand{\vp}{\varphi}

\newcommand{\beqn}{\begin{equation}}
\newcommand{\eeqn}{\end{equation}}
\newcommand{\bea}{\begin{eqnarray}}
\newcommand{\eea}{\end{eqnarray}}
\newcommand{\beastar}{\begin{eqnarray*}}
\newcommand{\eeastar}{\end{eqnarray*}}
\newcommand{\bdpm}{\begin{displaymath}}
\newcommand{\edpm}{\end{displaymath}}

\newcommand{\refRef} [1]{\cite{#1}}

\newcommand{\refeq}  [1]{(\ref{#1})}

\newcommand{\refsect}[1]{section \ref{#1}}

\newcommand{\refsects}[2]{sections \ref{#1} and \ref{#2}}

\usepackage{amsmath,amssymb}
\usepackage{layout}
\usepackage{graphicx}

\begin{document}
\begin{titlepage}
\bigskip
\rightline{}
\rightline{hep-th/0210303}
\bigskip\bigskip\bigskip\bigskip
\centerline{\Large \bf {When Black Holes Meet Kaluza-Klein Bubbles}}
     \bigskip\bigskip
           \bigskip\bigskip

  \centerline{\large Henriette Elvang and Gary T. Horowitz}
      \bigskip\bigskip
  \centerline{\em Department of Physics, UCSB, Santa Barbara, CA 93106}
   \centerline{elvang@physics.ucsb.edu, gary@physics.ucsb.edu}
                \bigskip\bigskip

\begin{abstract}
We explore the physical consequences of a recently 
discovered class
of exact solutions to five dimensional Kaluza-Klein theory.
We find a number of 
surprising features including: (1)
In the presence of a Kaluza-Klein bubble, 
there are arbitrarily large black holes
with topology $S^3$. (2) In the presence of a black hole or a black
string, there are expanding
bubbles (with de Sitter geometry) which never reach null infinity. (3)
A bubble can hold two black holes of arbitrary size in static
equilibrium. In particular,
two large black holes can be close together without merging to form a single
black hole. 

\end{abstract}
\end{titlepage}

 \baselineskip=16pt

%%%%%%%%%%%%%%%%%%%%%%%%%%%%%%%%%%%%%%%%%%%%%%%%%%%%%%%
%% Title etc - end
%%%%%%%%%%%%%%%%%%%%%%%%%%%%%%%%%%%%%%%%%%%%%%%%%%%%%%%

%%%%%%%%%%%%%%%%%%%%%%%%%%%%%%%%%%%%%%%%%%%%%%%%%%%%%%%
%% INTRODUCTION
%%%%%%%%%%%%%%%%%%%%%%%%%%%%%%%%%%%%%%%%%%%%%%%%%%%%%%%
 
 \setcounter{equation}{0}
\section{Introduction}

In four dimensions, there is a well known static solution describing
two black holes held apart by a strut (conical singularity) running
between them \cite{Israel,Myers}. 
It is natural to ask if there is an analogous solution in higher dimensions.
One problem is that if one tries in higher dimensions to construct a
one dimensional strut using
a line density of matter, it is likely to form a horizon around it and
become a black string. A black string connecting two black holes is not
likely to be static, but will instead collapse to a single black hole.

In five dimensional Kaluza-Klein (KK) theory, this is one of the questions
that has recently been settled by a new class of exact solutions 
presented by Emparan and Reall \refRef{Emparan:2001wk}. There are,
indeed, static solutions involving two black holes which are
nonsingular everywhere outside the horizons. 
The new solutions were found by generalizing an approach first used by
Weyl \refRef{Weyl} to construct all static, axisymmetric vacuum solutions in 
four dimensions. In this case, one can choose suitable variables
so that Einstein's equation essentially 
reduces to a linear equation. Emparan and Reall showed that Weyl's approach
can be generalized to higher dimensional solutions with enough symmetry.

In the static two black hole solution, the role of the strut is played by
a Kaluza-Klein ``bubble of nothing". This bubble was first found by Witten
\refRef{Witten:gj} almost twenty years ago as the endstate of the decay of the KK
vacuum. It can be obtained by a double analytic continuation of the five
dimensional Schwarzschild solution.
The characteristic feature of the bubble is that the KK circle smoothly
shrinks to 
zero size at a finite radius. The result is a minimal area $S^2$ which is
called the bubble, or ``bubble of nothing" since there is no space inside.
In Witten's example, the bubble rapidly expands outward  and hits null
infinity.  Its intrinsic geometry is de Sitter space.

The new solutions describe various configurations of black holes and KK 
bubbles.
We explore the physical consequences of some of these solutions and find
a number of surprising features\footnote{The special case of just one 
black hole on a
KK bubble was examined in detail in \refRef{Emparan:2001wk} in terms of
C-metric coordinates. The possibility of two black holes on a bubble was
briefly mentioned in \refRef{Emparan:2001wk}, but the solution
was not explicitly constructed.}.
For example, one might have thought that a black hole with $S^3$
topology in KK theory is possible only if its size is smaller than
the size of the circle at infinity. A much larger black hole should become a 
black string with topology $S^2 \times S^1$. Nevertheless, it turns out that
in the presence of a KK bubble, there are arbitrarily large black holes
with $S^3$ topology. Furthermore, two of these large black holes
can be held in static equilibrium by just a small piece of bubble!
Normally one expects that when two
black holes are brought together, they merge into one; another event horizon
usually forms which encloses
both. However, we find that  even when the separation between the
black holes is a negligible
fraction of their size, the solution remains static and does not form a
single black hole. Yet another surprising feature
involves the behavior of expanding
bubbles in the presence of a black hole or black string.
We will see that there are
bubbles whose intrinsic geometry is de Sitter space, yet the bubbles
never reach null infinity.

These results should be contrasted with related results about black holes
and bubbles in Kaluza-Klein theory.
It was shown in \cite{Myers} that arbitrarily large
$S^3$ black holes can exist for a fixed size circle at infinity.
However, that was for extremally charged black holes in
a  five dimensional Einstein-Maxwell theory. In that case, the horizon
remains a round sphere, and the size of the circle grows as one comes in
from infinity. We are considering
neutral black holes and five dimensional vacuum solutions. We will see
that the circle does not grow, but the horizon can distort itself
to fit into the space.
It has also been shown that bubbles constructed from the Kerr metric 
do not reach null infinity 
\cite{Aharony:2002cx}.
However, our examples appear to be the first de Sitter bubbles 
with a complete null infinity.

Since KK bubbles can arise from the quantum instability of the vacuum,
one would expect to produce not just one, but several expanding
bubbles. The question 
then arises: what happens when two bubbles collide. 
In a recent paper \refRef{Horowitz:2002cx}, colliding Kaluza-Klein
bubbles are studied in a 3+1-dimensional spacetime. The relevant
solution   is obtained from the previously mentioned two black hole
solution in four dimensions by analytic continuation. 
The resulting metric has no struts and it describes two colliding
$S^1$ Kaluza-Klein bubbles. It is shown that the collision produces a
black hole with a Schwarzschild type singularity inside. 

It was argued in \refRef{Horowitz:2002cx} that a similar bubble collision
in 4+1 dimensions should have symmetry $SO(2,1) \times U(1)$, and hence
not be a Weyl solution. What happens if one analytically continues
the five-dimensional static two black hole solution? We find a new solution 
describing two $S^2$ KK bubbles stuck on a (pre-existing) black string.
The bubbles collide inside the horizon, but this is simply because 
everything inside the horizon must hit 
the singularity.

It was suggested in \refRef{Emparan:2001wk} that  by continuously
changing parameters one could find examples of
black holes turning into black strings.  
This would be extremely interesting, since the nature of the transition
between static black holes and black strings is not well understood
(see, e.g., \cite{Kol:2002xz}). However,
when examined more closely we find that this  transition occurs in the
space of Weyl solutions only with a drastic change in the boundary conditions
at infinity. If one keeps the boundary conditions fixed, continuous changes
of the parameters do not change the horizon topology.

We begin with a
brief review of the four and  five dimensional Weyl
solutions describing simple black holes and bubbles. In \refsect{s-2BHKK} we
present and analyze the exact solution describing two black holes on a
KK bubble. The physical consequences are discussed in
\refsect{s-physcon}, including the demonstration that a small piece of
bubble can hold two large black holes apart.  
In \refsect{s-analytic} we consider analytic continuations
of the metric of \refsect{s-2BHKK}. In \refsect{s-collide} we analyze
the solution describing two bubbles on a black string and  show that
there are expanding bubbles with de Sitter geometry which do not reach null
infinity. In \refsect{s-3KK} we present a
different analytic continuation describing 3 adjacent Kaluza-Klein
bubbles. Finally, we discuss our results in \refsect{s-Disc}.

%%%%%%%%%%%%%%%%%%%%%%%%%%%%%%%%%%%%%%%%%%%%%%%%%%%%%%%
%% BLACK HOLE SOLN REVIEW
%%%%%%%%%%%%%%%%%%%%%%%%%%%%%%%%%%%%%%%%%%%%%%%%%%%%%%%

 \setcounter{equation}{0}
\section{Review of the Weyl Solutions}
\label{s-review}

We begin by reviewing the Weyl form of simple black hole and bubble solutions
in four and five dimensions. These will form the basis for the more general
solutions we discuss later.

\subsection{Four Dimensions}
Static, axisymmetric vacuum solutions in four dimensions can
be written in the form
\bea \label{gen4D}
  ds^2 = - e^{2U} dt^2 
         + e^{-2U} r^2 d\phi^2       
         + e^{2\nu} \left( dr^2 + dz^2 \right) \, ,
\eea
where $U$ is an axisymmetric solution to the Laplace
equation in the flat metric $dr^2 + r^2 d\phi^2 + dz^2$,
and $\nu$ is determined by
$U$. 

If $U$ is the Newtonian potential of a rod of mass $M$
and length $2M$, the metric \refeq{gen4D} describes a Schwarzschild
black hole of mass $M$. Note that even though the effective source for $U$ is
only axisymmetric, the resulting spacetime is spherically symmetric.
If one analytically continues $t\to i\chi$
one obtains the euclidean black hole. Regularity requires  that
$\chi$ be periodically identified so one has Kaluza-Klein boundary
conditions. If one further analytically continues $\phi \to i\tau$,
one obtains the four dimensional Kaluza-Klein bubble.  The bubble is
defined to be the surface where $g_{\chi\chi}=0$, but this is
just the analytic continuation of the $S^2$ that was the black hole
horizon. This means that the induced metric on the bubble is two dimensional
de Sitter space. Thus, even though the
metric looks static, the bubble is really expanding.
The form of the metric one obtains from
\refeq{gen4D} corresponds to writing de Sitter space in static
coordinates.

If $U$ is chosen to be the Newtonian potential of two 
non-intersecting rods with masses $M_i$ and lengths $2M_i$, $i=1,2$,
the metric \refeq{gen4D} describes two black holes. The
black holes are held apart by forces due to struts that arise from
conical singularities on the axis between the two black holes. The
configuration requires a single strut between the two black holes or
alternatively two struts extending from each black hole to infinity
along the axis of symmetry. This construction generalizes to 
solutions of $n$ collinear black holes: the lengths of the rods
determine the sizes of the black holes, and in general there will be
struts between the black holes.

% If $U_1$ is the potential of infinitely many identical rods at equal
% separation, the solution is regular and there are no need for struts
% (ref Israel-Kahn, Myers, ... (see ER paper for refs).

Starting with two equal size static black holes and
analytically continuing $t\to i\chi$ and  $\phi \to i\tau$, the
metric \refeq{gen4D} describes two expanding $S^1$ Kaluza-Klein
bubbles. This solution, which is completely free of conical deficits, was
analyzed in \refRef{Horowitz:2002cx}. 

%%%%%%%%%%%%%%%%%%%%%%%%%%%%%%%%%%%%%%%%%%%%%%%%%%%%%%%

\subsection{Five Dimensions}
As shown in \refRef{Emparan:2001wk}, the Weyl ansatz can be extended
to higher dimensions.  
Static, five dimensional vacuum solutions with two additional commuting
isometries can
be written in the form
\bea \label{gen5D}
  ds^2 = - e^{2U_1} dt^2
         + e^{2U_2} d\phi^2  
         + e^{2U_3} d\psi^2       
         + e^{2\nu} \left( dr^2 + dz^2 \right) \, ,
\eea
where for $i=1,2,3$, the functions $U_i$ are axisymmetric solutions to
Laplace's equation in three-dimensional flat space such that  
$U_1 + U_2 + U_3 = \log r + \rom{constant}$. The function $\nu$ is
determined by the $U_i$'s. It is convenient to again take the $U_i$
to be the potentials of rods placed along the $z$ axis with mass per unit
length $1/2$. (Other choices tend to produce naked singularities.) For
each $i$,
$U_i=\log r$ near a rod, so the corresponding metric function always vanishes at
the rods. This produces event horizons when the associated coordinate is
timelike, or axes for rotational symmetry when the associated coordinate
is spacelike. If $U_1$ has no source, and $U_2$ has a rod of length $2M$,
one obtains the trivial product of time and the euclidean Schwarzschild
solution. This is the static $S^2$ Kaluza-Klein bubble.

Following \refRef{Emparan:2001wk}, let $U_1$ be the potential of a rod with mass
per length $1/2$ placed on the $z$-axis for $-\mu < z < \mu$, and
let $U_2$ and $U_3$ be the potentials of semi-infinite rods with 
$z > \mu$ and $z < -\mu$ 
respectively, and mass per length $1/2$. The solution
\refeq{gen5D} then describes a five-dimensional Schwarzschild black hole
with Schwarzschild radius $2\sqrt{\mu}$. 
 By analytically continuing $t \to i\chi$ and $\phi \to i\tau$
 the solution becomes that 
of an expanding Kaluza-Klein bubble which describes the decay
\refRef{Witten:gj} of the Kaluza-Klein vacuum $M^{3,1} \times S^1$.
As before,
the metric induced on the bubble is three dimensional de Sitter space ---
the analytic continuation of the round metric on the black hole horizon.

In analogy to the four-dimensional case, there are static solutions
involving two five-dimensional black holes. As we shall see
in the next section,
in five dimensions the role of a strut providing the force to hold the
two black holes apart is played by a Kaluza-Klein bubble; however, no 
conical deficits are associated with the bubble and even for black
holes of different sizes we find that the solution is free of conical
singularities.

%%%%%%%%%%%%%%%%%%%%%%%%%%%%%%%%%%%%%%%%%%%%%%%%%%%%%%%
%% TWO BHS ON KK BUBBLE
%%%%%%%%%%%%%%%%%%%%%%%%%%%%%%%%%%%%%%%%%%%%%%%%%%%%%%%

 \setcounter{equation}{0}
\section{Two Black Holes on a Kaluza-Klein Bubble}
\label{s-2BHKK}

In \refsect{s-solution}
we construct the solution describing two black holes
on a Kaluza-Klein bubble. 
We identify the horizons and the bubble in
\refsect{s-horizons}, and we analyze the asymptotic behavior 
in \refsect{s-asymp}.

\subsection{The Solution}
\label{s-solution}
We study the metric \refeq{gen5D}
where $U_i$, $i=1,2,3$, are the Newtonian potentials of line-masses
with densities $1/2$ such that for positive real numbers $a$, $b$, $c$
with $a,c >b$ we have sources (see Fig.\ 1)
\begin{figure}[t]
  \begin{center}
      \includegraphics[width=8.4cm]{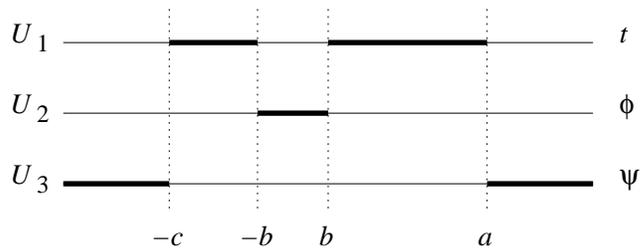}
        \end{center}
  \caption{The bold lines denote the effective sources along the $z$-axis
  for each of the three functions $U_1, U_2, U_3$. The right hand side 
  shows the coordinates associated with each $U_i$.}
	  \end{figure}
\begin{itemize}
  \item[$U_1$:] Two finite rods positioned on the $z$-axis for
$-c<z<-b$ and $b<z<a$;
  \item[$U_2$:] One finite rod positioned on the $z$-axis for
$-b<z<b$; 
   \item[$U_3$:] Two semi-infinite rods positioned on the $z$-axis for 
$z<-c$ and $z>a$.
\end{itemize}
Define as in \refRef{Emparan:2001wk} 
\bea
  \zeta_1 = z-a ~  ;~~~
  \zeta_2 = z-b ~  ;~~~
  \zeta_3 = z+b ~  ;~~~
  \zeta_4 = z+c \, ,
\eea
and for $i=1,\dots,4$
\bea
  R_i    &=& \sqrt{r^2 + \zeta_i^2} \\
  Y_{ij} &=& R_i R_j + \zeta_i \zeta_j + r^2 \, .
\eea
We can then write the first three metric components as
\bea 
  \label{U1}
  g_{tt} \; = \; - e^{2U_1} &=& 
               - \frac{(R_2 - \zeta_2)(R_4 - \zeta_4)}
                    {(R_1 - \zeta_1)(R_3 - \zeta_3)} \\
  \label{U2}
  g_{\phi\phi} \;  = \; e^{2U_2} &=& 
                   \frac{R_3 - \zeta_3}{R_2 - \zeta_2} \\ 
  \label{U3}
  g_{\psi\psi} \;  = \; e^{2U_3} &=& \left(R_1 - \zeta_1 \right)
                        \left(R_4 + \zeta_4 \right) \, ,
\eea
and solving for $\nu$ using the methods described in
\refRef{Emparan:2001wk}, we find
\bea \label{nu}
  g_{rr} \; = \; g_{zz} \; = \; e^{2\nu} &=& 
               \frac{Y_{14} Y_{23}}{4 R_1 R_2 R_3 R_4}
               \sqrt{ \frac{Y_{12} Y_{34}}{Y_{13} Y_{24}}}
               \left( \frac{R_1 - \zeta_1}{R_4 - \zeta_4} \right) \, .
\eea
%%%%%%%%%%%%%%%%%%%%%%%%%%%%%%%%%%%%%%%%%%%%%%%%%%%%%%%
%% REGULARITY
%%%%%%%%%%%%%%%%%%%%%%%%%%%%%%%%%%%%%%%%%%%%%%%%%%%%%%%
%\subsection{Regularity}
The periodicity of $\psi$ and $\phi$ are fixed by the requirement of 
regularity on the associated axes of rotation. As mentioned above,
these axes correspond to the location of the rods for $U_2$ and $U_3$,
since $U_i \to -\infty$ near the rod causing the 
corresponding metric component to vanish. Thus
for $z<-c$ and $z>a$, the orbit of $\psi$ vanishes in the limit 
$r \to 0$. Regularity requires that in these $z$-regions 
\bea \label{regu}
   \lim_{r\to 0} 
    \frac{\sqrt{g_{\psi\psi}} \Delta\psi}
         {\int_{0}^r g_{rr} dr} =2\pi \, ,
\eea
where $\Delta\psi$ denotes the period of $\psi$.
For both cases, $z<-c$ and $z>a$, we find that \refeq{regu} is
satisfied by taking $\Delta\psi = 2\pi$.

For $|z| < b$, the orbit of $\phi$ vanishes as $r\to 0$.
Taking $\phi$ to have period
\bea \label{Dphi}
  \Delta \phi =  \frac{8 \pi b(a+c)}{\sqrt{(a+b)(b+c)}}  
\eea
the solution is regular and free of conical deficits.

%%%%%%%%%%%%%%%%%%%%%%%%%%%%%%%%%%%%%%%%%%%%%%%%%%%%%%%
%% HORIZONS
%%%%%%%%%%%%%%%%%%%%%%%%%%%%%%%%%%%%%%%%%%%%%%%%%%%%%%%

\subsection{Event Horizons and Bubbles}
\label{s-horizons}

The metric component $g_{tt}$ vanishes when $r = 0$ and $b<z<a$ or when  
$r = 0$ and $-c<z<-b$, corresponding to the two black hole horizons.  
For the first horizon the constant-$t$ metric with $r=0$ is given by 
\bea\label{bhgeom}
  ds_{\rom{bh1}}^2 
    =  \frac{z-b}{z+b}  ~d\phi^2
      +  4 (a-z)(z+c) \, d\psi^2 
      +  \frac{(a+c)^2 (a-b) \, (z+b)}
              {(a+b) \, (z-b) (a-z) (z+c)} ~dz^2
\eea
where $b<z<a$, and for the second horizon the metric is
\bea\label{bh2geom}
  ds_{\rom{bh2}}^2 
    =  \frac{z+b}{z-b}  ~d\phi^2
      +  4 (a-z)(z+c) \, d\psi^2 
      +  \frac{(a+c)^2 (c-b) \, (z-b)}
              {(b+c) \, (z+b) (a-z) (z+c)} ~dz^2
\eea
where $-c<z<-b$. Since the $\phi$ circles shrink to zero size at $z=b$ and the
$\psi$ circles shrink to zero size at $z=a$ or $z=-c$,
we see that topologically the
horizons are $S^3$. The area of the first horizon is
\bea \label{BHh1}
  A_\rom{bh1} & = & 
   \frac{32 \pi^2 b\, (a+c)^2  \, (a-b)^{3/2}}{(a+b) (b+c)^{1/2}}
\eea
and the area of the second horizon is obtained by interchanging $a$
and $c$ in \refeq{BHh1}. 

When we checked regularity we observed that for $|z|<b$, the orbit of
$\phi$ vanished for $r=0$. This means that between the two
black holes sits a Kaluza-Klein ``bubble of nothing''. The metric on
this minimal bubble, for constant $t$ and $r=0$, is given by
\bea \label{bubgeom}
  ds_\rom{bubble}^2 = 
     4 (a-z)(c+z) d\psi^2 
   + \frac{4 b^2 (a+c)^2 }{(a+b) (b+c)} \frac{dz^2}{b^2-z^2}  \, . 
\eea
Note that the $\psi$-orbit does not close off at $z = \pm b$, so that
the ``bubble'' is actually a cylinder rather than an $S^2$; yet 
we shall continue to refer to it as a bubble.

The proper distance between the two black holes is
\bea \label{propdist}
  s = \frac{2 \pi b (a+c) }{\sqrt{(a+b)(b+c)}}
\eea
along a curve of constant $\psi$.

%%%%%%%%%%%%%%%%%%%%%%%%%%%%%%%%%%%%%%%%%%%%%%%%%%%%%%%
%% ASYMPTOTICS
%%%%%%%%%%%%%%%%%%%%%%%%%%%%%%%%%%%%%%%%%%%%%%%%%%%%%%%

\subsection{Asymptotic Behavior and Total Mass}
\label{s-asymp}

Consider to leading order the asymptotic behavior of the
metric. As $\rho = \sqrt{r^2 + z^2} \to \infty$, we find to order 
$O(\rho^{-2})$
\bea
  g_{tt} &=& -\left(1 - \frac{a+c-2b}{\rho}\right) \\ 
  g_{\phi\phi} &=& 1-\frac{2b}{\rho} \\
  g_{\psi\psi} &=& r^2
                   \left( 1  
                          + \frac{a+c}{\rho} \right) \\
  g_{rr} ~ = ~  g_{zz} &=& 1+\frac{a+c}{\rho} \, .
\eea
The leading order metric is
\bea
  ds_{\rho \to \infty}^2 \sim 
     - dt^2 
     + d\phi^2
     + r^2 d\psi^2 + dr^2 + dz^2 \, ,
\eea
so asymptotically the space is $M^{3,1}\times S^1$. The $S^1$ at
infinity is parametrized by $\phi$ and its size is given by
\refeq{Dphi}. Notice that the size of the circle at infinity is
precisely four times the distance between the two black holes
\refeq{propdist}. Note also that $M^{3,1}$ has a complete null infinity;
the asymptotic flat metric is not cut-off.
Both of these facts will be important in the next section.

Using the next to leading order metric $h_{\mu\nu}$, we
compute the total mass of the configuration. 
The ADM mass is given by
\bea
  M &=& \frac{1}{16\pi} \lim_{\rho \to \infty}
        \int_{S^2 \times S^1} 
        \left( \p_i h_{ij} - \p_j h_{ii} \right) N^j dV \, ,
\eea
where the surface $S^2 \times S^1$ is at constant $\rho$;
here $N^i$ is the unit normal vector of this surface and indices 
$i$ and $j$ label coordinates $\phi$, $x$, $y$, and $z$ with
$x=r\cos\psi$ and $y=r\sin\psi$. We find
\bea \label{Mtotal}
  M &=& \frac{4 \pi b (a+c-b)(a+c) }{\sqrt{(a+b)(b+c)}} \, .
\eea
It is clear that the total mass is always positive.

%%%%%%%%%%%%%%%%%%%%%%%%%%%%%%%%%%%%%%%%%%%%%%%%%%%%%%%
%% PHYSICAL CONSEQUENCES
%%%%%%%%%%%%%%%%%%%%%%%%%%%%%%%%%%%%%%%%%%%%%%%%%%%%%%%

 \setcounter{equation}{0}
\section{Physical Consequences}
\label{s-physcon}

In this section we analyze the physical behavior of the solution. We
consider first the limit of two small black holes on a KK bubble
(\refsect{s-smallBHs}). We then let the black holes be much larger
than their separation, and also study what happens when
the separation goes to zero (\refsect{s-bigbhs}). It turns out that big
black holes on a KK bubble resemble fat black strings, so in
\refsect{s-entropy} we compare the entropy of a black string with that
of the two black hole solution. 

\subsection{Small Black Holes}
\label{s-smallBHs}

To gain physical intuition about this solution, we start with the
case $a=c=b+\ep$ where $\ep \ll b$. For $\ep=0$, it is easy to see
that $g_{tt} = -1$ and
the solution reduces to the product of time and the euclidean Schwarzschild
solution. This is
the static Kaluza-Klein bubble with topology $S^2$, radius $2b$, and circle at
infinity of length $8\pi b$.  For $\ep \ne 0$, one
adds two small black holes on opposite sides of this bubble. The geometry
of the black holes \refeq{bhgeom}, \refeq{bh2geom}
simplify considerably in the limit of small $\ep$.
Since the horizon corresponds to
$b\leq |z| \leq b +\ep$, $g_{\phi\phi} = O(\ep)$. So to leading order in 
$\ep$, we can ignore the change in the periodicity of $\phi$. Let
$\phi = 4b \vp$ where $\vp$ has period $2\pi$. Then to leading order in
$\ep$, the horizon geometry (for $z>0$) becomes
\bea
  ds_\rom{bh1}^2 = 8b(z-b)d\vp^2 + 8b(b+\ep -z) d\psi^2 + 
     \frac{2b\ep \ dz^2}{(z-b)(b+\ep-z)}
\eea
Letting $z=b+\ep\sin^2\theta$, with $0\leq \theta \leq \pi/2$, this
becomes 
\bea
   ds^2_\rom{bh1} = 
     8b\ep(d\theta^2 + \sin^2\theta d\vp^2 +\cos^2\theta d\psi^2)
\eea
The second black hole is similar.
So the black holes are round three spheres with radius $\sqrt{8b\ep}$.
Normally a small black hole in Kaluza-Klein theory would be localized on
the $S^1$, but that is not what is happening here. The solution retains
the rotational symmetry around the circle. A small spherical black hole
is possible because the size of the compact direction shrinks to zero
on the KK bubble.

%%because the KK bubble shrinks the size of the compact direction
%%to zero.

At first sight, it is surprising that one can add black holes to a
static bubble and keep the solution static. One might expect that
adding some mass would cause the bubble to collapse. One can
understand what is happening as follows. For a fixed size circle at
infinity, one can construct a one parameter family of initial data
describing bubbles of varying radii 
\refRef{Brill:qe}. 
The initial data is time symmetric and  the metric takes the form
\bea
   ds^2 = F(\rho) d\chi^2 + F^{-1}(\rho) d\rho^2 + \rho^2 d\Omega
\eea
where $F(\rho) = 1-4M/\rho -c/\rho^2$. The two free parameters are the
total mass 
$M$, and an arbitrary constant $c$.
The bubble is located at the positive zero of 
$F$, $\rho = \rho_+ \equiv 2M+\sqrt{4M^2+c}$. 
Regularity requires that $\chi$ be periodic with period
\bea\label{inidata}
   L= \frac{2\pi \rho_+^2}{\rho_+ - 2M} \, .
\eea
This is the length of the circle at infinity.
Although the full evolution of this initial data is not known, Corley and
Jacobson \refRef{Corley:mc} showed that the second derivative of the
bubble area with respect to proper time along the bubble is given by 
\bea
  \frac{d^2 A}{d\tau^2} = 8\pi\left (1- \frac{4M}{\rho_+}\right )
\eea
Using \refeq{inidata} this can be rewritten as
\bea
  \frac{d^2 A}{d\tau^2} =8\pi\left (\frac{4\pi \rho_+}{L} -1\right )
\eea
This clearly shows that there is a linear relationship between the bubble size
and its initial acceleration. 

Now consider the metric on the bubble \refeq{bubgeom}.
If one starts with the static bubble $a=c=b$, and increases $a$ and $c$ to
add the two black holes,
it is easy to see that the bubble size increases. So the
interpretation is that the new bubble would normally accelerate outward, but is
kept in place by the black holes. Conversely, the attraction between the
black holes is exactly  canceled by the  natural expansion of the bubble.
Since the general static solution has $a\ne c$, it is clear that the
bubble can adjust itself to support unequal mass black holes. The fact that
the solution remains static even with only one black hole on the bubble
($c=b, a>b$) indicates that the black hole can attract the bubble itself and
stop it from expanding. We will see a clear example of this in section 
(5.1).

Now let us continue to increase the size of the black holes.
One might expect that the black holes will merge when they become larger
than the size of the bubble. However, it is easy to see that this does
not happen. We have already remarked that the separation between the black 
holes is always one fourth of the size of the circle at infinity. So 
keeping our asymptotic metric fixed, the black holes never touch.
To see what happens we now turn to the opposite limit of black holes
much larger than the size of the bubble.

%%%%%%%%%%%%%%%%%%%%%%%%%%%%%%%%%%%%%%%%%%%%%%%%%%%%%%%

\subsection{Big Black Holes}
\label{s-bigbhs}

We now study the limit of two big black holes stabilized by the bubble
between them.  
Taking $a, c \gg b$ the bubble metric \refeq{bubgeom}
reduces to 
\bea\label{thinbub}
  ds^2_\rom{bubble} = 4ac \, d\psi^2
                    + \frac{4 b^2(a+c)^2}{ac} \, d\eta^2 \,  
\eea
where $z = - b \cos\eta$ with $0 < \eta < \pi$. The bubble is a flat
cylinder with large radius $2\sqrt{ac}$ and small height 
$2\pi b(a+c)/\sqrt{ac}$.

Away from the bubble, 
the metric of each black hole horizon is approximately
\bea\label{bigbh}
  ds^2_\rom{bh1,2} = d\phi^2 + 4(a-z)(z+c) \, d\psi^2 
                    + \frac{(a+c)^2}{(a-z)(z+c)} \, dz^2
\eea
where $b \ll z < a$ for horizon 1 and $-c < z  \ll -b$ for horizon
2. This can be further simplified by defining 
$z=[(a-c)+(a+c)\cos\theta]/2$. Then the horizon metric becomes
\bea
ds^2_\rom{bh1,2} = d\phi^2 + (a+c)^2[d\theta^2 + \sin^2\theta d\psi^2]
\eea
This is just the product of a circle and (part of) a round two-sphere
of radius $a+c$. The first black hole corresponds to roughly
$0\leq \theta < \theta_0$
where $\cos\theta_0 \approx (c-a)/(a+c)$, while the second is
approximately 
$\theta_0 < \theta \leq \pi$. 
The estimate for $\theta_0$ is
only approximate since when $|z|$ approaches $b$, the horizon no
longer takes the form \refeq{bigbh}. In this region, the radius of
the $\phi$ circles becomes $z$ dependent and shrinks to zero at $|z| = b$.
This is how one obtains arbitrarily large black holes with $S^3$ topology
in a space  with one direction compactified. One starts with a large
four dimensional black hole, which in Kaluza-Klein theory is really a 
$S^2\times S^1$ black string. One then adds a KK bubble which pinches off the
horizon changing its topology to $S^3$.

\begin{figure}[t]
  \begin{center}
    \includegraphics[width=5.5cm]{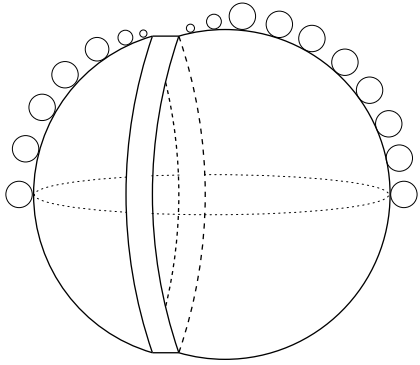}
  \end{center}
  \caption{Geometry of a thin cylindrical Kaluza-Klein bubble separating
    two large $S^3$ black holes. One can view this as
    starting with a black string $S^2\times S^1$ (with the $S^2$ radius
    much larger than the circle), cutting the $S^2$ into two pieces, separating
    the pieces, and gluing in a KK bubble.}
\end{figure}

The solution analyzed here
provides a way of deforming a black string with horizon 
$S^2\times S^1$ into a configuration of two large $S^3$ black holes, 
separated by a
cylindrical KK bubble (see Fig.\ 2). 
If the radius of the black string $S^2$ is much
larger than the radius of the $S^1$, we can cut the $S^2$ along
\emph{any} latitude and insert a KK bubble between the two parts such
that the $\phi$-orbit on the horizon shrinks smoothly to zero where
the bubble intersects the horizon. 
In \refsect{s-entropy}, we compare the entropy of the original black 
string to the entropy of the black holes on the bubble.  

It seems quite remarkable that a small piece of Kaluza-Klein bubble
can hold two large black holes apart. Not only does it compensate for
the expected gravitational attraction, but it also prevents the black
holes from merging. In $3+1$ dimensions, if one brings two black holes
close together, a third horizon forms surrounding the two. (This can
be clearly seen in terms of initial data \cite{Cook:gp}.) In these Kaluza-Klein
solutions, this does not happen even when the separation between the
black holes is negligible compared to their size. We do not have an
intuitive explanation for this. It is easy to see that adding the bubble
does not change the total mass significantly: From \refeq{Mtotal} and
\refeq{Dphi}, one sees that the total mass divided by the length of the
circle at infinity is $M/L = (a+c-b)/2 \approx (a+c)/2$. This agrees with
the mass per unit length of a black string with Schwarzschild radius $a+c$
and length $L$.

One possibly relevant fact is that
if one includes the time direction, the metric on the bubble
\refeq{thinbub} becomes  
\bea
  ds^2_\rom{bubble} = 4ac \, d\psi^2
                    + \frac{4 b^2 (a+c)^2 }{ac} \, \left[d\eta^2 - 
		    \frac{\sin^2\eta}{4(a+c)^2} dt^2\right] \, ,
\eea
Thus the bubble is the 
product of a circle, and a static patch of two 
dimensional de Sitter spacetime.
We have seen that the coordinate $t$ is a
unit time 
translation at infinity. The form of $g_{tt}$ means
that there is a redshift  between the bubble and infinity which can be
interpreted as arising from the proximity of the black holes.
The de Sitter geometry indicates that
the bubble is expanding, but only a static patch
appears outside the black holes. So the distance between the black holes
remains constant. The de Sitter horizons coincide with the
black hole horizons. 

What happens to the bubble inside the black holes?
The bubble is defined by $g_{\phi\phi}=0$. The black hole horizon is a 
three sphere, and $g_{\phi\phi}=0$ just selects a circle on this sphere.
The usual Penrose diagram for a black hole suppresses the entire sphere,
so it corresponds to fixing one point on the sphere. If one fixes this
point to lie on the line $g_{\phi\phi}=0$, then the bubble extends
throughout the black hole in a Penrose diagram. On the other side of the
black hole throat is another region of spacetime which is identical 
to the one on this side. So there is another static patch of a de Sitter 
bubble on the other side which joins onto the first. The result is shown
in Fig.\ 3.\footnote{We have made the simplest assumption that the 
spacetime on the other side of the two black holes is the same.}
\begin{figure}[t]
  \begin{center}
    \includegraphics[width=7cm]{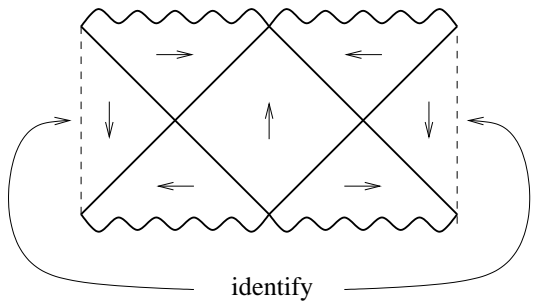}
  \end{center}
  \caption{A Penrose diagram of the bubble with the $\psi$ direction
  suppressed. The bubble extends through
  the two black holes. In the limit of large black holes, the geometry of the
  bubble outside the black holes is a static patch of de Sitter. The 
  arrows denote the $\p/\p t$ symmetry.}
\end{figure}
The global topology of the bubble is now $T^2$ since the direction through
the black holes provides a second circle in addition to the one parameterized
by $\psi$.
This qualitative discussion can be made more precise using the coordinates
introduced in section 5, which extend the metric inside the event horizons
and beyond the black hole throat.

Most other two dimensional slices through the geometry have a more
conventional description. For instance, if we fix
$\phi$, $\psi$ and $z$, the 
geometry on the $r,t$ plane depends on the value of $z$. If $z>a$ or $z<-c$,
then the spacetime is causally the same as two dimensional Minkowski space.
If $b<z<a$ or $-c<z<-b$, the causal structure is the same as Schwarzschild,
with its  two asymptotically flat regions.
If $-b<z<b$, and we now
include the $\phi$ circles, the geometry looks like
a static bubble: the radius of the $\phi$ circles goes from a constant at
infinity to zero on the bubble. There is a redshift between the bubble and
infinity which increases as $|z|$ approaches $b$. 

As we discussed in section 2, the Weyl solutions can be characterized
by the rods which act like sources for the functions $U_i$.
 From a comparison of the configurations of rods for the two black holes
on a Kaluza-Klein bubble with that of a four-dimensional Schwarzschild
times a flat direction --- i.e.\ a black string --- it is tempting to
conclude that the limit $b\to 0$ corresponds to a transition between
black holes and a black string. This would be very interesting since
one expects (in the absence of KK bubbles) that a small $S^3$
black hole in Kaluza-Klein theory becomes a black string when its size 
increases, yet exactly how this transition occurs is not understood.
Unfortunately, the rods are misleading. 
It is true that in the
limit $b \to 0$, the length of Kaluza-Klein bubble goes to zero
and the two black holes approach each other. However the size of
the circle at infinity also vanishes. In fact, in this limit,
the two horizons match up to
form a perfect two-dimensional sphere and the solution becomes that of a 
four-dimensional Schwarzschild black hole with a four-dimensional mass
$(a+c)/2$.  However, this limit is very singular, since e.g.\ the curvature
on the Kaluza-Klein bubble diverges as its size goes to zero.
It appears one cannot use these Weyl metrics to study the 
black hole -- black string transition.

%%%%%%%%%%%%%%%%%%%%%%%%%%%%%%%%%%%%%%%%%%%%%%%%%%%%%%%

\subsection{Entropy}
\label{s-entropy}

In a spacetime which asymptotically is $M^{3,1} \times S^1$ we have
studied the exact solution describing two black holes sitting on a
Kaluza-Klein bubble. In terms of the parameters $a$, $b$, and $c$
the circle at infinity has length
\bea \label{theL}
  L = \frac{8 \pi b(a+c)}{\sqrt{(a+b)(b+c)}} \, ,
\eea
the ADM mass of the configuration is
\bea \label{theM}
  M = \frac{4 \pi b(a+c-b)(a+c) }{\sqrt{(a+b)(b+c)}} \, ,
\eea 
and the total area of the two black holes is
\bea \label{theArea}
  A_\rom{2BH} = A_\rom{bh1} + A_\rom{bh2}
   = \frac{32 \pi^2 b(a+c)^2 }{(a+b)^{1/2} \, (b+c)^{1/2}}
     \left[   \frac{(a-b)^{3/2}}{(a+b)^{1/2}}
            + \frac{(c-b)^{3/2}}{(c+b)^{1/2}}
     \right] 
\eea

Since the black holes resemble a black string everywhere away from the 
bubble,
we compare this area to the area of a five-dimensional black string
with the same mass \refeq{theM} and size of circle at infinity \refeq{theL}.
 The black string metric can be written 
\bea
  ds^2_\rom{BS} = - \left(1 - \frac{R_0}{R}\right) dt^2
         + \left(1 - \frac{R_0}{R}\right)^{-1} dR^2
         + R^2 \, d\Omega_2^2
         + dz^2
\eea
with $z \sim z + L$. The ADM mass of the black string is $M = R_0 L/2$
and the horizon area is $A_\rom{BS} = 4\pi R_0^2 L$. Thus in 
terms of the mass, the area is $A_\rom{BS} = 16 \pi M^2/L$. Inserting
the values \refeq{theL} and \refeq{theM} we find, using dimensionless
scalings $x, y > 1$ defined by $a = xb$ and $c = yb$, that the ratio
of the areas is given by 
\bea
  \frac{A_\rom{2BH}}{A_\rom{BS}} =
    \frac{x+y}{(x+y-1)^2}
      \left[ \frac{(x-1)^{3/2}}{(x+1)^{1/2}} 
             + \frac{(y-1)^{3/2}}{(y+1)^{1/2}}\right] \, .
\eea
Analyzing this function we find that for all $x,y>1$,
\bea
  A_\rom{2BH} < A_\rom{BS}
\eea
so that the simple black string is always entropically favored over the
configuration of two black holes on a Kaluza-Klein bubble. Hence we
should not expect a black string to spontaneously generate a KK bubble
that splits the black string horizon $S^1 \times S^2$ into two black
hole $S^3$ horizons connected by the bubble.

%%%%%%%%%%%%%%%%%%%%%%%%%%%%%%%%%%%%%%%%%%%%%%%%%%%%%%%
%% ANALYTIC CONTINUATION
%%%%%%%%%%%%%%%%%%%%%%%%%%%%%%%%%%%%%%%%%%%%%%%%%%%%%%%

 \setcounter{equation}{0}
\section{Analytic Continuation}
\label{s-analytic}

In this section we consider double analytic continuations of the metric
describing two black holes on a Kaluza-Klein bubble. We shall do this  
in two ways, leading to two distinct solutions: one describing two
$S^2$ KK bubbles on a black string 
(\refsect{s-collide}), the other describing three adjacent
KK bubbles (\refsect{s-3KK}). The first solution has the standard
Kaluza-Klein boundary conditions, but the second does not --- it
approaches $M^3\times S^1\times S^1$ asymptotically. For this reason,
we concentrate on the first solution and discuss the second only
briefly. 

%%%%%%%%%%%%%%%%%%%%%%%%%%%%%%%%%%%%%%%%%%%%%%%%%%%%%%%
%% COLLIDING BUBBLES
%%%%%%%%%%%%%%%%%%%%%%%%%%%%%%%%%%%%%%%%%%%%%%%%%%%%%%%

\subsection{Two bubbles on a black string}
\label{s-collide}

We analytically continue the solution from \refsect{s-2BHKK} by taking
$t \to i\chi$ and $\phi \to i\tau$, so that the metric is now
\bea
  ds^2 =   e^{2U_1} d\chi^2 
         - e^{2U_2} d\tau^2
         + e^{2U_3} d\psi^2
         + e^{2\nu} \left( dr^2 + dz^2 \right) \, ,
\eea
with $U_i$ and $\nu$ given by equations \refeq{U1}-\refeq{nu}. To avoid
conical singularities, we must set 
$a=c$ and make $\chi$ 
periodic with period  
\bea \label{deltachi}
  \Delta \chi = 8 \pi a \sqrt{\frac{a-b}{a+b}} \, .
\eea
The Kaluza-Klein circle at infinity is now parameterized by $\chi$.

When $r \to 0$, $g_{\tau\tau}$ vanishes for $|z|<b$, so at $r=0$ and
$|z|<b$ we have a horizon. This horizon has topology $S^2\times S^1$,
so it is a black string. 
The $S^2$ is parameterized by $z$ and $\chi$, and the $S^1$ is
parameterized by $\psi$. 
The constant-$\tau$ metric of the horizon is
\bea
  ds_\rom{horizon}^2 = \frac{b^2 - z^2}{a^2 - z^2} d\chi^2
      + 4\, (a^2 - z^2)\, d\psi^2
      + \frac{16 \, a^2 \, b^2}{(a+b)^2} \frac{dz^2}{b^2 - z^2}
  \, ,
\eea
with $|z|<b$.
The area of the horizon is 
\bea
  A_\rom{horizon} = \frac{(16 \pi a b)^2}{a+b} 
    \sqrt{\frac{a-b}{a+b}} \, .
\eea

For $b<|z|<a$, the orbit of $\chi$ vanishes as $r \to 0$. This means
that at $r=0$ we have two KK bubbles with $b<z<a$ and $-a<z<-b$,
respectively. The metric of the first bubble is (constant $\tau$ and
$r=0$) 
\bea \label{disk}
  ds_\rom{bubble1}^2 = 4\, (a^2 - z^2) \, d\psi^2
      + \frac{4 \, a^2 \, (a-b)}{a+b}
        \left( \frac{z+b}{z-b} \right)
        \frac{dz^2}{a^2 - z^2} 
\eea
with $b<z<a$. Topologically, this describes a disk not an $S^2$ since
the orbit of $\psi$ does not close off at $z=b$. However, these
coordinates do not cover the entire constant-$\tau$ surface. Consider a
geodesic with constant $\tau, \chi, \psi$ as it approaches $r=0$.
If either $g_{\chi\chi}$ or $g_{\psi\psi}$ vanishes at $r=0$, then this is
just the axis of a rotational symmetry, and the geodesic continues to positive
values of $r$ with $\chi$ or $\psi$ shifted by  half its period. This is
the case for $|z|\geq b$. However, for $|z|<b$, both $g_{\chi\chi}$ and
$g_{\psi\psi}$ remain nonzero and one can continue the spacetime past $r=0$.
Since the metric only depends on $r^2$, the natural extension is to let
$r$ become negative. This yields another copy of the geometry which is exactly
analogous to the region on the other side of the Schwarzschild throat in
the maximally extended Schwarzschild geometry.
The spacetime is clearly
invariant under $r \to -r$  which implies that the black string
horizon at $r=0, |z| \leq b$
is a minimal $S^2\times S^1$. 

Returning to the bubble, we now see that there is another copy of the
disk \refeq{disk} on the negative $r$ side and the two disks smoothly
join at $z=b$ to make an $S^2$. A similar argument applies to the bubble
at $z=-b$. The net result is that the configuration on  a constant $\tau$
slice describes two $S^2$ bubbles on opposite sides of a $S^2\times S^1$
horizon. The spacetime appears static, but since the coordinates do not
cover the entire spacetime, this is misleading. We now show that under
evolution the two bubbles collide at the black string singularity.

When one considers the evolution of this spacetime, one immediately faces an
apparent contradiction. Consider the limit when $a$ is close to $b$, so the
bubbles are small. We saw in the previous section, that before the analytic
continuation, this spacetime described two small
$S^3$ black holes on a KK bubble. Since the black holes are round spheres,
after the analytic continuation, the bubble geometry must be three dimensional
de Sitter space $dS_3$. In other words, these are expanding bubbles. 
It was shown in \refRef{Aharony:2002cx} that de Sitter bubbles expand
out and hit null infinity. 
However, it is clear from the asymptotic form of the metric that null
infinity is complete. So the bubble never reaches null infinity.
The resolution is that the bubbles are held in by the black string. Most
of each bubble lies inside the black string, and only the static patch
extends outside the horizon (see Fig.\ 4).
\begin{figure}[t]
  \begin{center}
      \includegraphics[width=6.4cm]{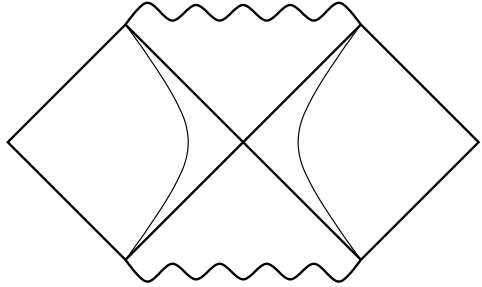}
   \end{center}
	  \caption{The evolution of an $S^2$ bubble sitting on an
      $S^2 \times S^1$ black string. The bubble geometry is de
      Sitter, but does not reach null infinity. There is another
      bubble on the opposite side of the $S^2 \times S^1$ black string
      which is not shown.} 
\end{figure}

As we proceed with the analysis of the collision of the two bubbles,
we shall follow the analysis of \refRef{Horowitz:2002cx} closely. 
In order to describe the spacetime to the future of the black string
horizon, we analytically continue $r$, taking $\tilde{r} = ir$.\footnote{In
this analytic continuation, $e^{2\nu}$ must be continuous, and hence stay
positive.} Then
$\tilde{r}$ is the time coordinate and $\tau$ becomes spacelike. We
introduce double null coordinates  
\bea
  u = \tilde{r} + z, ~~~~\rom{and}~~~~ v = \tilde{r} - z 
\eea
and just as in \refRef{Horowitz:2002cx} we find that the metric in terms of $u$
and $v$ has coordinate singularities on each of the null hypersurfaces
$u=b$ or $v=b$. In terms of a better set of null coordinates 
\bea
  U = - \sqrt{b-u}  ~~~~\rom{and}~~~~  V = - \sqrt{b-v}
\eea
the metric is no longer singular at $u=b$ or $v=b$. However, along 
\bea\label{UVcircle}
  U^2 + V^2 = 2b 
\eea
some metric components diverge. We now show that this is just a coordinate
singularity along the two 
 timelike segments, $UV<0$, corresponding to the motion of the bubbles.
 However, along the 
future spacelike segment, $U, V >0$, there is a real curvature singularity.
The past spacelike segment, $U, V <0$, is simply the black string horizon
encountered above. 

We now establish these claims. We expand the metric near the time-like
segment of \refeq{UVcircle} along the null hypersurface $V = V_0 < 0$, taking 
\bea \label{U-coord}
  U = \sqrt{2b - V_0^2} - \ep \, ,
\eea
where $\ep > 0$ is assumed to be small.  To leading order
the metric behaves as
\bea
  g_{\chi\chi} \sim O(\ep^2) \, ,
  ~~~~
   g_{\tau\tau} \sim O(1) \, ,
  ~~~~
  g_{\psi\psi} \sim O(1) \, ,
  ~~~~
   g_{UV} \sim O(1)  \, .
\eea
This corresponds to a flat space four-dimensional geometry times a circle
parametrized by $\psi$. Thus the metric remains nonsingular while the
bubbles approach each other.

For the future spacelike segment, $V = V_0 > 0$ and $U$ again given by
\refeq{U-coord}, we find
\bea \label{BS-sing}
  g_{\chi\chi} \sim O(\ep^4) \, ,
  ~~~~
   g_{\tau\tau} \sim O(\ep^{-2}) \, ,
  ~~~~
  g_{\psi\psi} \sim O(1) \, ,
  ~~~~
   g_{UV} \sim O(\ep^4)  \, .
\eea
This is like a black string singularity. To see this, note that the
metric for the five-dimensional Schwarzschild black string near the
singularity can be written
\bea
  ds_\rom{BS}^2 \approx \frac{2m}{r} d\tau^2
     - \frac{r}{2m} dr^2 
     + r^2 (d\theta^2 + \sin^2{\theta} d\chi^2) 
     + d\psi^2
\eea
Setting $r = \ep^2$ and introducing double null coordinates
\refRef{Horowitz:2002cx}, $U = \sqrt{2/m} \, \ep + \theta$ and 
$V = \sqrt{2/m} \, \ep - \theta$, we obtain the same behavior as in
\refeq{BS-sing}.

We conclude that the bubbles evolve from the $r=0$ surface, and as
they approach each other the metric remains non-singular. The
bubbles collide at the null points $(U,V) = (\sqrt{2b},0)$ and 
$(U,V) = (0,\sqrt{2b})$ corresponding to the black string curvature
singularity. These points are at null proper distance because 
$g_{UV} \to 0$.

The question remains whether the black string is formed by the collision
of the two bubbles (as in \cite{Horowitz:2002cx})
or whether the bubbles are just 
sitting on opposite
poles of the $S^2$ on the $S^2\times S^1$ horizon
and brought together simply because the entire $S^2$ shrinks to a point
at the singularity. The correct interpretation is the latter.
A simple collision of two KK bubbles should have symmetry $SO(2,1) \times
U(1)$, since a single bubble has symmetry $SO(3,1) \times U(1)$, and
a second bubble will break the Lorentz group down to the subgroup acting
orthogonal to the direction between the bubbles. This symmetry is
incompatible with the Weyl ansatz. By adding the black string, one replaces
the $SO(2,1)$ symmetry by a time translation and axisymmetry. This can be seen
by looking at the induced metric at $z=0$. This is the surface exactly
half-way between the bubbles. In the simple colliding bubble solution,
the $\psi $ circles should shrink to zero size, and the metric should have
$SO(2,1)$ symmetry. The fact that the $\psi $ orbit is nonzero on the
horizon indicates that there is a pre-existing black string.

%%%%%%%%%%%%%%%%%%%%%%%%%%%%%%%%%%%%%%%%%%%%%%%%%%%%%%%
%% Other Analytic Cont
%%%%%%%%%%%%%%%%%%%%%%%%%%%%%%%%%%%%%%%%%%%%%%%%%%%%%%%
\subsection{Other Configurations}
\label{s-3KK}

Consider first $c=b$. Before analytic continuation the metric describes
a single black hole sitting on a Kaluza-Klein bubble. 
Making a double analytic
continuation of this metric, taking $t \to i\chi$ and $\phi \to
i\tau$, interchanges the roles of the black hole and bubble. If the
original configuration described a small black hole sitting on a big
bubble ($a-b \ll b$; as in \refsect{s-smallBHs}, but with $a \ne c$
and $c = b$), the solution obtained by analytic continuation
describes a big black hole on a tiny bubble ($a-b \gg b$; as in 
\refsect{s-bigbhs} with $c=b$). 
The coordinates do not describe the full constant-$\tau$ surface: there
is another asymptotic region described by letting $r$ take negative
values. In this region, we find (by symmetry in $r \to -r$) the other
''half'' of the Kaluza-Klein bubble, so that in the full space the
bubble is topologically an $S^2$. The solution of a black hole on a KK
bubble was studied in \refRef{Emparan:2001wk} in terms of C-metric
coordinates.

Now let $b \ne c$.
As in \refsect{s-collide}, we analytically continue the metric
with $a=c$ by taking $t \to i\chi$, but this time we
take $\psi \to i\tau$. Again regularity requires $\chi$ to be periodic
with period \refeq{deltachi}.
This configuration describes three adjacent KK bubbles at $r=0$:
for $|z| < b$ we have an $S^2$ bubble parametrized by $z$ and $\chi$,
and for $b<|z|<a$ we find two bubbles parametrized by $z$ and
$\phi$. For $r=0$ and $|z| > a$ we encounter Rindler horizons.
Extending the $r$-coordinate to run over all real values, we find that
the two latter bubbles are also topologically $S^2$.
The solution has two $S^1$'s at infinity, so asymptotically it is 
$M^3\times S^1\times S^1$ and not the Kaluza-Klein vacuum. 
Although we have not studied this solution in detail, it appears that
at least two of the bubbles expand outward and hit null infinity. It is
thus remarkable that the bubbles appear
never to collide and the spacetime remains nonsingular.

%%%%%%%%%%%%%%%%%%%%%%%%%%%%%%%%%%%%%%%%%%%%%%%%%%%%%%%
%% DISCUSSION
%%%%%%%%%%%%%%%%%%%%%%%%%%%%%%%%%%%%%%%%%%%%%%%%%%%%%%%

 \setcounter{equation}{0}
\section{Discussion}
\label{s-Disc}

Higher dimensional gravity plays an important role in recent discussions of
string theory, M-theory, and brane worlds. We have seen that it has many 
unexpected properties. Using the methods of \cite{Emparan:2001wk} we have
constructed and studied a three parameter family of exact solutions
describing the interaction of black holes and Kaluza-Klein bubbles.
One of the most surprising results is that a small piece of bubble  can
support two 
enormous black holes in static equilibrium. The gravitational attraction 
of the black holes is apparently balanced by the tendency of the bubble
to expand. 
In the spirit of the principle of maximum tension recently discussed
by Gibbons \refRef{Gibbons:2002iv}, we find it interesting that we
have a static solution with arbitrarily close black holes supported
by a KK bubble. However, in this limit it is difficult to define
the force between the black holes, so we cannot discuss whether
there is a
corresponding bound on a maximum repulsive force.

%For equal size black holes, their four dimensional
%radius is of order $a$ while their separation is of order $b$, so 
%a rough  Newtonian estimate
%of the force between them is $F\sim a^2/b^2$ which can be arbitrarily large.
%It is remarkable that Kaluza-Klein bubbles can be so strong. 
%In a recent paper \refRef{Gibbons:2002iv}, Gibbons motivates a
%principle of maximum tension. Our examples indicate that there
%cannot be a corresponding bound on a maximum repulsive force.

One might hope that these solutions allow one to study the
transition between horizon topologies, for example two black holes
merging to form a black string in KK theory. As we have seen, the Weyl
solutions we have studied here are not suitable for this purpose. 
The fact that the proper distance \refeq{propdist} between the black
holes is one fourth of the size of the circle at infinity \refeq{Dphi}
implies that we cannot let the black holes approach each other without
shrinking the circle at infinity. In
\refsect{s-bigbhs} we saw that in the limit $b \to 0$, the circle at
infinity vanishes and the black hole horizons join up to form the
round $S^2$ horizon of a four-dimensional Schwarzschild black hole. 
However, the curvature blows up on the KK bubble, so the limit is
singular. 

The relation between the black hole separation and the size of the circle at 
infinity is a consequence of the requirement of regularity for the
$\phi$-coordinate in 
the region $|z|<b$ when $r \to 0$. We can ease up on this requirement
at the cost of introducing conical singularities, and it is natural to
ask how this will influence the black hole merger.
Now instead of fixing the period of $\phi$ by
\refeq{Dphi} as required by regularity,
let us set $\Delta \phi = 2\pi k$ for some positive constant $k$. 
This introduces a conical singularity at $r=0$ for $|z| < b$, and
including the $\psi$-direction we find that the conical singularity
is actually spread over the surface of the KK bubble. The deficit
angle associated with the conical singularity is
\bea
  \delta = 
   2\pi \left( 1- \frac{k \sqrt{(a+b)(b+c)}}{4b (a+c)}\right) \, . 
\eea
Note that for given $a$, $b$, and $c$ the angle $\delta$ can be
positive or negative, depending on the choice of $k$. If 
$k < 4b(a+c)/\sqrt{(a+b)(b+c)}$, the angle is a deficit angle so the
strut provides a pull. However, since the period of $\phi$ is smaller
than in \refeq{Dphi}, the black holes are now smaller than in the
configuration without the strut. So the bubble now
balances the smaller black holes as well as the pull of the strut. If 
$k > 4b(a+c)/\sqrt{(a+b)(b+c)}$, the angle is an excess angle. The
combined efforts of the bubble and the strut can now balance two
bigger black holes.

Now let the black holes approach each other by taking $b \to 0$. If $k
\propto b$, then $\Delta \phi \to 0$ as $b \to 0$, so the circle at
infinity vanishes. Alternatively, if $k$ approaches some nonzero
constant as $b \to 0$, then for sufficiently small $b$ the angle
$\delta$ is an excess angle and in the limit $b \to 0$ this
excess angle diverges.
We conclude, as in \refsect{s-bigbhs}, that even in the presence of
conical singularities the rod picture is
deceptive: when $b \to 0$, the rod picture suggests that the two
black holes merge to form a four-dimensional Schwarzschild black 
hole times a flat direction --- a black string --- however, when
taking regularity and struts into account we have shown that this
limit is singular.

In \refsect{s-bigbhs} we showed how a fat black string can be deformed
by cutting the $S^2$ along any latitude, inserting a piece of KK
bubble, and requiring the $S^1$ to smoothly shrink to zero size where
the bubble intersects the horizons. The result is two black holes
separated by a KK bubble. Away from the bubble, the horizon of a
big black hole again  looks like a black string and we can repeat the
cutting and gluing process to obtain a solution
describing two black holes and one black string 
held apart by KK bubbles. This is illustrated in Fig.\ 5. If we
analytically continue $t \to i\chi$ and $\phi \to i\tau$ we obtain a
solution with two long thin black strings separated by a KK bubble,
and on each black string sits an $S^2$ KK bubble. 
Generalizing the above, we find solutions describing $n$
collinear black strings separated by KK bubbles; at each end we either
have a black hole with an $S^3$ horizon or an $S^2$ KK bubble. 
Thus solutions with multiple black strings and black holes
combine the configurations studied in \refsects{s-physcon}{s-analytic}. 

\begin{figure}[t]
  \begin{center}
    \raisebox{1.1cm}{
      \includegraphics[width=7.8cm]{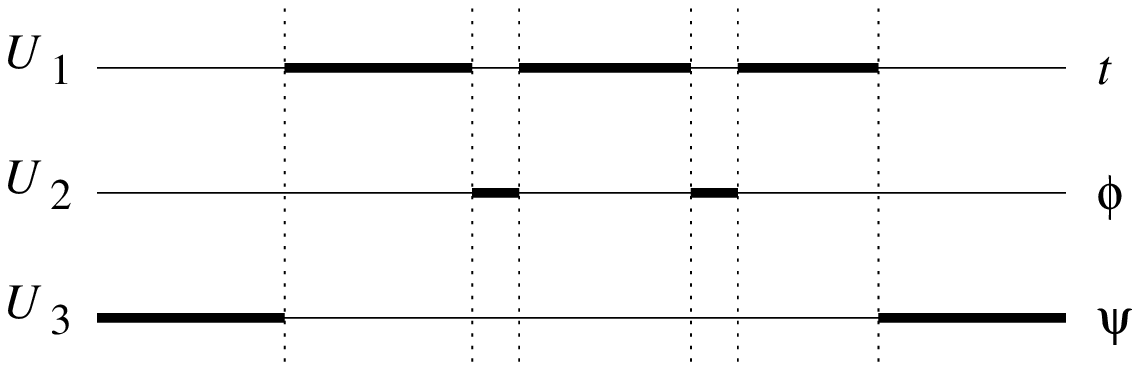}}
      ~~~~
      \includegraphics[width=6cm]{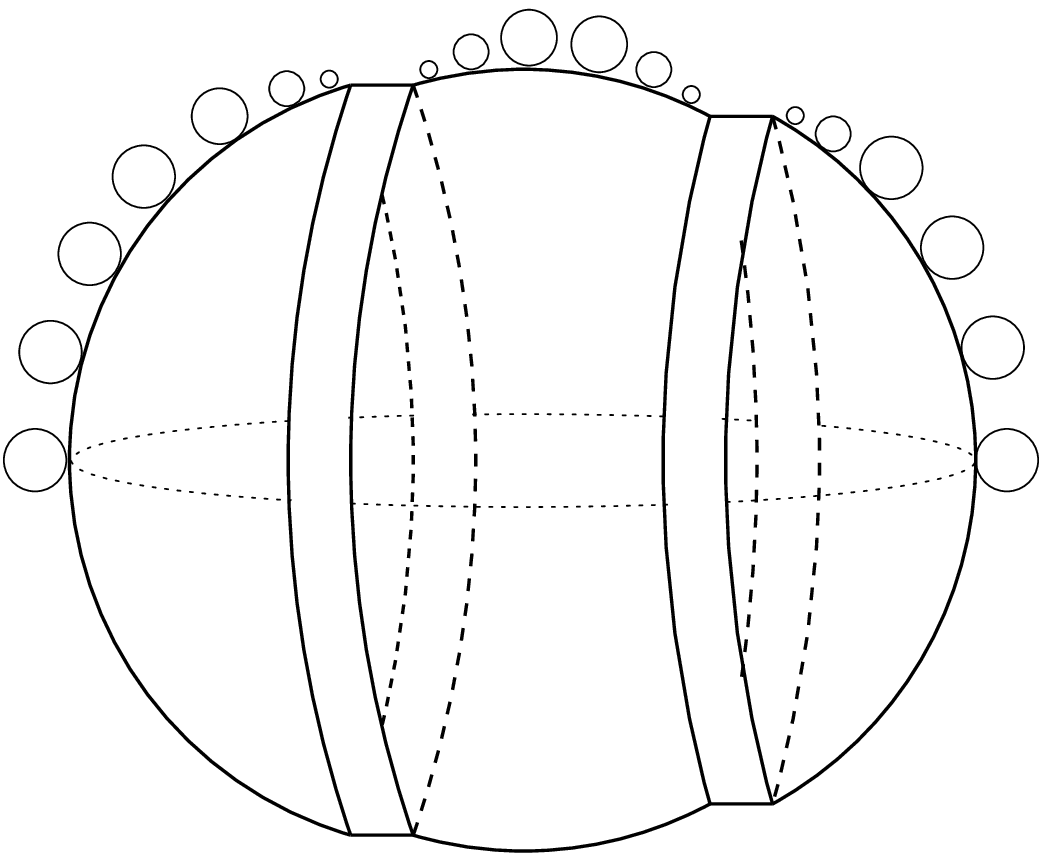}
   \end{center}
	  \caption{Rod configuration for the solution of two $S^3$
    black holes and a black string held apart by KK bubbles. To avoid
    conical singularities, the rod sources for the $U_2$ potential have
    to have the same length. On the sketch of the configuration, the
    circle over each point is parametrized by $\phi$. For the black
    string in the middle, $\phi$ is the angular coordinate of the
    $S^2$ and the $S^1$ is parametrized by $\psi$.}
\end{figure}

We have not investigated the stability of the solutions. On the one
hand, one might argue that
the solutions describing two static black holes on a bubble should be unstable
since they represent a delicate balance between the attraction of the
black holes and expansion of the bubble. Intuitively,
if a perturbation increases the
distance between the black holes, their attraction will decrease, but the
expansion may increase, causing the black holes to accelerate apart.
Conversely, a perturbation which brings the black holes slightly closer may
cause the system to collapse into a single black hole. On the other hand,
this Newtonian argument might fail for large black holes supported by a
bubble. In this case, the bubble geometry is de Sitter space
and is expected to be stable
\cite{Aharony:2002cx}. 
For the solution in 5.1 describing
two KK bubbles on a black string,  in
the limit $a\gg b$ the black string takes the form of a small
round $S^2$ times a long $S^1$. This is precisely the regime in which 
one usually expects the black string to be unstable \cite{Gregory:vy}.
Can the large KK bubbles stabilize the black string?

We may at this point speculate about generalizations of our solution
to higher dimensions. These cannot be achieved as a Weyl
solution: the $d$-dimensional generalized Weyl solutions have $d-2$
commuting Killing vector fields, which are not permitted by
$d>5$-dimensional Schwarzschild solutions. So the interaction 
of higher dimensional black holes and bubbles will require a new
class of solutions.

\section*{Acknowledgement}

We would like to thank R. Emparan,  A.~Maharana and E. Martinec for discussions.
This work was supported in part by NSF grant PHY-0070895 and  the
Danish Research Agency.

%%%%%%%%%%%%%%%%%%%%%%%%%%%%%%%%%%%%%%%%%%%%%%%%%%%%%%%
%% REFS
%%%%%%%%%%%%%%%%%%%%%%%%%%%%%%%%%%%%%%%%%%%%%%%%%%%%%%%


\begin{thebibliography}{}
\bibitem{Israel}
W.~Israel and K.~Khan, ``Collinear Particles and Bondi Dipoles in
General Relativity", Nuovo Cimento, {\bf 33} 331 (1964).
\bibitem{Myers}
R.~Myers, ``Higher Dimensional Black Holes
In Compactified Space-Times''
Phys.~Rev.~{\bf D35}, 455 (1987).

%\cite{Emparan:2001wk}
\bibitem{Emparan:2001wk}
R.~Emparan and H.~S.~Reall,
``Generalized Weyl solutions,''
Phys.\ Rev.\ D {\bf 65}, 084025 (2002)
[arXiv:hep-th/0110258].
%%CITATION = HEP-TH 0110258;%%
\bibitem{Weyl} H.~Weyl, Ann.~Phys.~(Leipzig) {\bf 54} (1917) 117.
%\cite{Witten:gj}
\bibitem{Witten:gj}
E.~Witten,
``Instability Of The Kaluza-Klein Vacuum,''
Nucl.\ Phys.\ B {\bf 195}, 481 (1982).
%%CITATION = NUPHA,B195,481;%%


%\cite{Aharony:2002cx}
\bibitem{Aharony:2002cx}
O.~Aharony, M.~Fabinger, G.~T.~Horowitz and E.~Silverstein,
``Clean time-dependent string backgrounds from bubble baths,''
JHEP {\bf 0207}, 007 (2002)
[arXiv:hep-th/0204158].
%%CITATION = HEP-TH 0204158;%%

%\cite{Horowitz:2002cx}
\bibitem{Horowitz:2002cx}
G.~T.~Horowitz and K.~Maeda,
``Colliding Kaluza-Klein bubbles,''
arXiv:hep-th/0207270.
%%CITATION = HEP-TH 0207270;%%

%\cite{Kol:2002xz}
\bibitem{Kol:2002xz}
B.~Kol,
``Topology change in general relativity and the black-hole black-string  transition,''
arXiv:hep-th/0206220.
%%CITATION = HEP-TH 0206220;%%


%\cite{Brill:qe}
\bibitem{Brill:qe}
D.~Brill and G.~T.~Horowitz,
``Negative Energy In String Theory,''
Phys.\ Lett.\ B {\bf 262}, 437 (1991).
%%CITATION = PHLTA,B262,437;%%
%\cite{Corley:mc}
\bibitem{Corley:mc}
S.~Corley and T.~Jacobson,
``Collapse Of Kaluza-Klein Bubbles,''
Phys.\ Rev.\ D {\bf 49}, 6261 (1994)
[arXiv:gr-qc/9403017].
%%CITATION = GR-QC 9403017;%%

%\cite{Cook:gp}
\bibitem{Cook:gp}
G.~B.~Cook and A.~M.~Abrahams,
``Horizon Structure Of Initial Data Sets For Axisymmetric Two Black Hole Collisions,''
Phys.\ Rev.\ D {\bf 46}, 702 (1992).
%%CITATION = PHRVA,D46,702;%%

%\cite{Gibbons:2002iv}
\bibitem{Gibbons:2002iv}
G.~W.~Gibbons,
``The maximum tension principle in general relativity,''
arXiv:hep-th/0210109.
%%CITATION = HEP-TH 0210109;%%

%\cite{Gregory:vy}
\bibitem{Gregory:vy}
R.~Gregory and R.~Laflamme,
``Black Strings And P-Branes Are Unstable,''
Phys.\ Rev.\ Lett.\  {\bf 70}, 2837 (1993)
[arXiv:hep-th/9301052].
%%CITATION = HEP-TH 9301052;%%



\end{thebibliography}
\end{document}